\documentclass[sigconf]{acmart}
\usepackage{graphicx} 
\usepackage{comment}
\usepackage{hyperref}
\usepackage{tabularx}
\usepackage{booktabs}
\usepackage{enumitem}

\copyrightyear{2024}
\acmYear{2024}
\setcopyright{rightsretained}
\acmConference[IDE '24]{2024 First IDE Workshop}{April 20, 2024}{Lisbon, Portugal}
\acmBooktitle{2024 First IDE Workshop (IDE '24), April 20, 2024, Lisbon, Portugal}\acmDOI{10.1145/3643796.3648447}
\acmISBN{979-8-4007-0580-9/24/04}

\begin{document}

\title{Help Me to Understand this Commit! -\\ A Vision for Contextualized Code Reviews}

\author{Michael Unterkalmsteiner}
\orcid{0000-0003-4118-0952}
\affiliation{
    \institution{Blekinge Institute of Technology, Software Engineering Research Lab}
    \city{Karlskrona}
    \country{Sweden}
}
\email{michael.unterkalmsteiner@bth.se}

\author{Deepika Badampudi}
\orcid{0000-0002-6215-1774}
\affiliation{
    \institution{Blekinge Institute of Technology, Software Engineering Research Lab}
    \city{Karlskrona}
    \country{Sweden}
}
\email{deepika.badampudi@bth.se}

\author{Ricardo Britto}
\orcid{0000-0002-7220-9570}
\affiliation{
    \institution{Ericsson AB and Blekinge Institute of Technology, Software Engineering Research Lab}
    \city{Karlskrona}
    \country{Sweden}
}
\email{ricardo.britto@ericsson.com}

\author{Nauman bin Ali}
\orcid{0000-0001-7266-5632}
\affiliation{
    \institution{Blekinge Institute of Technology, Software Engineering Research Lab}
    \city{Karlskrona}
    \country{Sweden}
}
\email{nauman.ali@bth.se}  

\begin{abstract}
    \textit{Background:} Modern Code Review (MCR) is a key component for delivering high-quality software and sharing knowledge among developers. Effective reviews require an in-depth understanding of the code and demand from the reviewers to contextualize the change from different perspectives. 
    \textit{Aim:} While there is a plethora of research on solutions that support developers to understand changed code, we have observed that many provide only narrow, specialized insights and very few aggregate information in a meaningful manner. Therefore, we aim to provide a vision of improving code understanding in MCR. 
    \textit{Method:} We classified 53 research papers suggesting proposals to improve MCR code understanding. We use this classification, the needs expressed by code reviewers from previous research, and the information we have \emph{not} found in the literature for extrapolation. 
    \textit{Results:} We identified four major types of support systems and suggest an environment for contextualized code reviews. Furthermore, we illustrate with a set of scenarios how such an environment would improve the effectiveness of code reviews.
    \textit{Conclusions:} Current research focuses mostly on providing narrow support for developers. We outline a vision for how MCR can be improved by using context and reducing the cognitive load on developers. We hope our vision can foster future advancements in development environments.  
\end{abstract}



\maketitle

\section{Introduction}
Modern Code Review (MCR) contributes to the code quality of software products~\cite{bavota2015} and knowledge dissemination in development organizations~\cite{bosu2016process}. The practice of MCR was born out of the need to keep up with the ever-increasing speed of development and release cycles~\cite{Bacchelli2013}, replacing formal code inspections that were introduced by Fagan in the 1970s~\cite{fagan1976} and which have seen a peak in research in the late 1990s~\cite{kollanus2009survey}. 
The MCR process relies on support that facilitates code distribution, feedback collection, and deciding whether to accept or reject a code change. The technological stack co-evolved with the MCR process comprises distributed version control systems (e.g., Git) and code collaboration tools (e.g., Gerrit and GitHub). The integration of these technologies has led to efficient management of the MCR process. 
To review code effectively, reviewers need to fully \emph{understand} the code changes with all their implications. The current technology stack, including source code repositories, management systems, and integrated development environments, does not sufficiently support this. The most common approach to reviewing a change is to compare old and new code without any additional information beyond the description the author may or may not have provided. 

Studies investigating code reviews in practice~\cite{Bacchelli2013, macleod2017code, ebert2021exploratory,soderberg2022understanding} show that one of the main challenges experienced by code reviewers is access to the necessary information to make effective review decisions. 
These studies indicate that the situation has not considerably changed during the past ten years. In a recent study from 2022, Söderberg et al.~\cite{soderberg2022understanding} identified a misalignment between the \emph{unit of analysis} that the review tools provide and the \emph{unit of attention} that a code reviewer utilizes to make a decision. With a unit of analysis, they refer to the files automatically compiled by the review tool (such as Gerrit or GitHub) based on the changed code. On the other hand, the unit of attention is the information a reviewer needs to make an informed decision. A key finding of their study is that, while a review starts with the unit of analysis, reviewers often need to go beyond the affected files to understand the overall code change. In a recent survey~\cite{badampudi2023modern}, practitioners rated \emph{understanding code under review} as the top research priority among existing MCR proposals. 

In this paper, we provide an overview of the existing proposals to help reviewers understand a code change, i.e., proposals expanding the unit of analysis to approximate the needed unit of attention. We have grouped these proposals into four categories: refactoring, change impact, code reorganization, and data integration. However, few, if any, of these ideas have been transferred to the technology stack that is used in industrial practice today. 

We argue that a fundamental problem of the identified solution proposals is their specialization on a single aspect of code understanding. They expand the unit of attention into one particular dimension only, for example, by decomposing significant changes into logical units. Other information needs are not satisfied by that particular solution proposal, and the unit of analysis and unit of attention remain misaligned. Without taking away the contributions that have been made in the past, we deem it necessary to rethink development environments for code reviews to increase their usefulness in practice.  
We envision a code collaboration environment whose main feature is providing context-dependent information, derived and synthesized from various artifacts and sources throughout the software development life-cycle and delivered on time. 


\section{Background}\label{sec:back}

\begin{figure}
    \centering
    \includegraphics[width=0.40\textwidth]{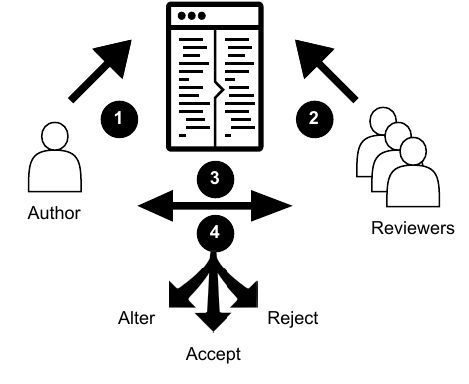}
    \caption{The code review process}
    \label{fig:crprocess}
\end{figure}

Generally, the modern code review process (Figure~\ref{fig:crprocess}) comprises the following four steps.
The code author prepares a code change (Step 1), typically together with a textual description of the purpose of the change. One or more reviewers (the selection of those is out of scope for our discussion) review the change. 
The change can include completely new or deleted or changed files, typically processed by a tool that visualizes and highlights the changes (Step 2). 
The reviewers provide written feedback on the change, possibly referring to concrete lines of code or providing a general assessment of the change. Change authors can respond to this feedback (Step 3). 
The outcome of the discussion is a decision on whether to alter the code and return to Step 1, accept or reject the change (Step 4). 

In this paper, we focus on Step 2, i.e., the information the reviewer(s) need to understand a change, evaluate it with respect to criteria that depend on the context of the review, and formulate actionable feedback for the change author.

\begin{table*}[!th]
    \caption{Code understanding categories used in review support systems}
    \label{tab:cats}
    \centering
    \footnotesize
    \begin{tabularx}{\textwidth}{lXl}
        \toprule
        Category & Definition & Papers \\
        \midrule
        Code reorganization & Code changes are reorganized (decomposed, reordered, regrouped) to allow for an easier understanding. & 26\\
        Data integration & The changed code is augmented with information, possibly in the form of visualizations, that provides additional context for the change. This information can but does not necessarily have to originate from the source code. & 11 \\
        Refactoring & The changed code is augmented with information about refactorings such that reviewers understand better the nature of the change. & 8 \\
        Change impact & Code changes are analyzed to understand how they impact related code (static analysis) or program behavior (dynamic analysis). & 8 \\
        \bottomrule
    \end{tabularx}
\end{table*}

\subsection{Related work}
Code reviewers' struggle with understanding code changes is not a new problem and has also been identified in traditional code inspections~\cite{dunsmore2000}. Bacchelli and Bird~\cite{Bacchelli2013} reported in 2013, in one of the first studies of MCR, that understanding the reasons for a change is the most complex and time-consuming aspect of the review. Reviewers read the change description, run the code, or even reach out to authors to understand the code. Furthermore, tool support is limited to providing visual differences in the code, syntax highlighting, and commenting. A key factor in providing quick and valuable feedback to the change author is a priori knowledge of the change context~\cite{Bacchelli2013}. 

Dong et al.~\cite{dong2021survey} found in their survey, which covered scientific and gray literature between 2006 and 2020, that understanding code context is still the most challenging aspect for code reviewers. Baum and Schneider~\cite{baum2016need} called in 2016 for a new generation of code review tools that trade in the flexibility of the supported review process with support for code change understanding for the reviewer. 

We argue that the trade-off posed by Baum and Schneider is not correct since the flexibility to support different review processes \emph{and} the adaptive fulfillment of reviewers' information needs should be a core feature of a modern code collaboration/review environment. Our view is also supported by the results of Pascarella et al.~\cite{pascarella2018information}, who studied the information needs of code reviewers. They conclude that the reviewers' information needs are on different conceptual levels and belong to different aspects of the code under review. Hence, the flexibility of the code review environment in terms of being able to satisfy information needs is essential.

To conclude, while context has been discussed many times (e.g., \cite{Bacchelli2013, pascarella2018information, dong2021survey, ebert2021exploratory} as being essential for improving code change understanding, the idea to provide contextualized supporting information on-demand and tailored to the current review information needs has, to the best of our knowledge, not yet been proposed. This paper fills this gap.

\section{Research Methodology}

Our starting point was the following research question: \emph{What aspects of code understanding for reviews are implemented in support systems?} To answer this question, we capitalized the results from a recent mapping study on MCR~\cite{badampudi2023modern}. We focused on 23 primary studies categorized as solution proposals to improve the understanding of the code under review. In March 2023, we conducted one iteration of forward snowball sampling~\cite{wohlin2014guidelines} on the start set, resulting in 351 papers. We identified 30 additional relevant papers. We read the full text of these 53 papers and extracted the investigated strategies to improve code understanding. The first author reviewed these strategies and identified four fundamental categories: refactoring, change impact, code reorganization, and data integration. Definitions of these categories are provided in Table~\ref{tab:cats}. The remaining authors reviewed the categorization. The data set containing the sampled papers and the categorization are available online~\cite{anonymous_2023}.


\paragraph{Validity threats}
We performed only a single iteration of forward snowball sampling, which was, however, based on a starting set (23 papers) derived from a recent and rigorously conducted mapping study~\cite{badampudi2023modern}. Another threat is that only the first author selected and categorized the studies from the snowball sample (30 papers), which may lead to the omission and misclassification of some studies. However, we reduced this threat by involving multiple researchers (study authors) in data extraction and developing the categories based on the 23 papers from the start-set. These stable categories and a shared understanding of the selection criteria helped to reduce the above threat.
While we acknowledge these threats, we consider that our approach was sufficient to take a broader overview of research, reflect on research trends and practitioners' needs (see Söderberg et al.~\cite{soderberg2022understanding}), and sufficiently demonstrate a research gap.  

\section{Code review contexts}\label{sec:contexts}
As part of our vision of the future of code reviews, we introduce the idea of \emph{code review contexts\footnote{Encyclop{\ae}dia Britannica defines to \emph{contextualize} as "to think about or provide information about the situation in which something happens" \url{https://www.britannica.com/dictionary/contextualize}}}. Murphy~\cite{murphy2019beyond} has argued that the lack of context in software engineering tools limits developers' effectiveness. Therefore, we envision a code review environment that provides reviewers with the required information about the code change based on the review context. 

The context can be global or local. Global means that the context is determined by the overall purpose of the review (e.g., identify security vulnerabilities) or the expertise of the reviewer (knowledge of the to-be-reviewed code). Local means that the context is determined by the file, class, method, or line, which is the current focus of the reviewer (unit of analysis). Depending on the global and local context, the reviewer has different information needs (unit of attention) that are satisfied by the envisioned code review environment, which we call Code Review Contextualizer (CoReCo) henceforth.
Using the lens of code review contexts, we reviewed the existing literature to identify the support systems that help reviewers understand code changes. This review aims to illustrate that many information needs can already be satisfied but must be \emph{integrated} into an environment that provides on-demand, contextualized support. 

In the remainder of this section, we provide an overview of solutions that have been proposed to improve code understanding in reviews (Section~\ref{sec:review}), our vision of a next-generation code review environment, CoReCo, (Section~\ref{sec:coreco}), and application scenarios that illustrate the potential of CoReCo (Section~\ref{sec:scenarios}).

\subsection{What we found}\label{sec:review} 

Table~\ref{tab:cats} describes the four code understanding categories we identified for review support systems. Systems with code reorganization are the most frequent proposals, with 26 studies.

\emph{Code reorganization.} 
Solutions are proposed to decompose composite change sets into different partitions based on (i) reasons for modifications, such as bug fixes and features, (ii) regrouping based on atomic changes addressing a single issue and related changes, and (iii) identifying essential parts of the commit. For example, SmartCommit~\cite{shen2021smartcommit}, a graph-partitioning-based interactive approach, is proposed to decompose tangled changesets automatically. An interactive approach called DiffSearch~\cite{di2022diffsearch} returns a set of changes matching a particular query. The queries are related to reasons for changing code, including code improvements and cleanup, functionality changes, bug fixes, and uses of a new API. Another approach is to help reviewers understand the salient class where the main change happens and triggers changes in the other classes. Huang et al.~\cite{huang2020code} propose a solution to characterize a class's salience based on several code and commit features.

\emph{Data integration.}
Providing a structural representation of the change can help reviewers understand the code to be reviewed. For example, Fregnan et al.~\cite{fregnan2023graph} propose a visualization approach that displays classes and methods in review changes as nodes in a graph. UTANGO~\cite{li2022utango} is another tool that decomposes different groups with different concerns within the changed code and its surrounding code. UTANGO visualizes additional contextual embeddings for code changes that integrate program dependencies, changes, contexts, and cloned code. 

\emph{Refactoring.}
Similar to the code reorganization category, the papers in this category reorganize the code with a focus on providing refactoring details. An example of such a solution is Refactorinsight~\cite{kurbatova2021refactorinsight}, a tool that auto-folds and provides a refactoring description, which in turn helps reviewers to focus on changes related to bug fixes and new features. Solutions are also proposed to identify specific refactorings, such as removing code clones~\cite{chen2018clone}.

\emph{Change impact.}
This category includes solutions that provide behavioral summaries describing the impact of changes and enable a better understanding of the code to be reviewed~\cite{menarini2017semantics}. Another example is a solution to provide relational information on the classes/methods/modules impacted by the change~\cite{mondal2021semantic}. 

\emph{Synthesis.}
The categories described in this section answer our initial research question, i.e., what aspects of code understanding for reviews are implemented in support systems? The proposals integrating data from different sources to provide context for the reviewing decision are the most interesting ones concerning aggregating information meaningfully, thereby reducing the misalignment between a unit of analysis and the unit of attention. Our proposal aims to develop this idea further,  explicitly with the goal of \emph{dynamic review contextualization based on the reviewers' needs}.

\subsection{The Code Review Contextualizer (CoReCo)}\label{sec:coreco}
\begin{figure}
    \centering
    \includegraphics[width=0.40\textwidth]{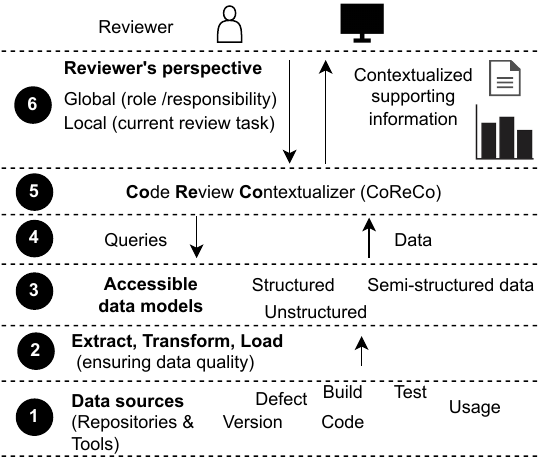}
    \caption{An overview of CoReCo's components}
    \label{fig:CoReCo}
\end{figure}

We envision a future where code reviews remain critical, even in a world where Large Language Models (LLMs) co-write most code. While generative tools help developers to accelerate production and explore alternatives~\cite{barke2023}, humans remain essential to ensuring high-quality code through reviews.

Human code reviewers must consider numerous context factors when reviewing code. The relationship between a code change and a product's architecture, security and safety constraints, performance, privacy, interoperability, maintainability (and many more) requirements may influence the reviewing strategy. An additional critical aspect of efficient code reviews is the order in which to approach them. A linear review may not be the most effective strategy, as critical changes in a commit can be located in the middle of a file. Managing these factors requires experience and is often challenging without tool support.
CoReCo is our vision of a code review environment that supports developers in making more efficient and effective review decisions as follows.

(1) CoReCo starts from the unit of analysis, i.e., the changed files, and analyzes its content, and, for example, change~\cite{mo2022} and defect~\cite{yu2020} history, frequency of execution in operation (e.g., through log mining techniques~\cite{el-masri2020}), and impact on the remaining system~\cite{kretsou2021}. By analyzing the content of changes and commit messages, CoReCo identifies (e.g., through topic modeling~\cite{silva2021}) relevant documentation from requirements specifications and informal communication. 
    
(2) CoReCo creates a profile of the unit of analysis with aspects (historically defect-prone code, mission-critical code, refactored code, etc.) that are likely important for the review. This profile, together with linked documentation, helps the reviewer to decide on what to focus on during the review. 

(3) Reviewer experience is characterized based on past reviews and used to estimate the alignment to the expertise requirements of the unit of analysis. Reviewing activities can be captured in real-time through data collection of interactions with CoReCo and biometrical data (e.g., detection of task-unrelated thoughts with webcams~\cite{hutt2023}). This allows CoReCo to track, in the long term, to what extent reviewer expertise and activities affect product quality.

(4) The interaction of the reviewer with CoReCo is primarily unidirectional, i.e., the system provides information consumed by the reviewer. However, we also foresee an interactive system (similar to Diffsearch~\cite{di2022diffsearch}, which returns information based on a query related to the code). Recent advancements in using LLMs for solving real-world problems, such as code explanation~\cite{leinonen2023comparing} and question-answering~\cite{petroni2019language}, are promising. Research on the potential of LLMs indicates their capability to comprehend code syntax rules but also their failure to comprehend (yet) dynamic program behavior~\cite{ma2023}.

Figure~\ref{fig:CoReCo} provides an overview of CoReCo's logical elements. CoReCo will sit on top of a system providing access to data collected throughout the software development life-cycle, from managing requirements, project planning, development, debugging environments and testing to product deployment and product usage (1). The data collection mechanism (2) will cope with diverse data sources and perform quality control. One key deterrent to using data analytics is poor data quality~\cite{figalist2022}. To make the data usable, data models must be developed (3) that allow clients, such as CoReCo, to access data on demand (4). Inspiration for such a data model can come from the work by Mart{\'i}nez-Fern{\'a}ndez et al.~\cite{martinez-fernandez2018} who propose an ontology for software engineering data integration. 
CoReCo prepares queries and receives data based on the review context (5). Such context intertwines the reviews' global and local perspectives. The global perspective is predominately determined by the reviewers' role, responsibility, and expertise. They set the initial context of the review and determine the initial parameters for CoReCo to adapt the queries and displayed data to the reviewers' information needs. During the review, CoReCo determines the context, i.e., the currently viewed code block, either by keyboard and mouse interaction or by tracking the reviewers' eye movement (or a combination of these two approaches). We expect that the work on code writing assistants, such as Copilot\footnote{\url{https://github.blog/2021-06-29-introducing-github-copilot-ai-pair-programmer}}, can be valuable here as they are also currently challenged to determine useful context~\cite{barke2023}. Finally, the information synthesized by CoReCo is displayed to the reviewers (6) when they need it. This can be a combination of a dashboard presenting information before the review is conducted, parameterized by the reviewers' perspective, and pop-up information screens that appear when relevant during the review. 

We envision CoReCo to support all the code understanding categories identified in Section~\ref{sec:review}, through the abstraction of data collection and extraction that renders the information accessible for upstream analysis tasks. We also foresee that this has the potential to boost the research in the data integration category, allowing the enrichment of review context with relevant information that is not easily accessible without such an infrastructure.



\subsection{Application Scenarios}\label{sec:scenarios}
To illustrate how we envision CoReCo being used in practice, we present a series of scenarios showing how our vision extends the work we reviewed (see Section~\ref{sec:review}). 

\emph{Security and deployment context.} An architect is reviewing code that belongs to an extensive distributed software system in the financial domain. In this case, we envision the architect getting relevant prompts from CoReCo when the code concerns authentication, authorization, or accounting (AAA). CoReCo will inform the architect where the code under review will be deployed and be accessible. Such information about AAA requirements and deployment plans will be beneficial for a reviewer to decide if the code has a security vulnerability. For example, if accounting is not a requirement, perhaps user management is not required as the code is part of a component that is deployed and accessible from only within the trusted network. 
CoReCo will process artifacts like the source code, the security policy, and deployment configurations.

\emph{High-risk code.} A developer is reviewing a pull request for a business-critical product. Failure to detect an issue in this module could lead to severe consequences since it is used frequently by most customers, as usage data shows. In the past, changes in this module have led to integration test failures, and impact analysis shows that files beyond the modified ones are affected. This information gives the reviewer indications that API changes in this module need to be carefully reviewed. Additional meta-data about the pull request, such as the familiarity of the author with the module based on past commits, and the time to the next planned release date, support the reviewer in deciding on what and why to comment.  

\emph{Code change history.} A developer is reviewing an unfamiliar pull request. The original code was authored several years ago and has seen four changes. Six months ago, the second change led to an intense discussion between the original author and the developer proposing the change. The second change was ultimately rejected. The third and fourth changes were minor and only addressed code style. The description of the change under review raises similar points as the code change from six months ago. While this information does not make the review task less challenging, the provided history on the controversial change provides context for the reviewer.

\section{Future plans}\label{sec:next}
We foresee three research themes that, while intertwined, could be investigated independently if the ultimate goal (transfer CoReCo to practice) and assumptions (that need to be verified throughout the research) are shared and agreed upon. 

\emph{Data collection, quality control, and accessibility.}
There is a need to design and implement infrastructure for collecting and validating software engineering data at scale from heterogeneous sources, which allows prototypical research solutions to be elevated to production environments~\cite{figalist2022}. The components 1-4 shown in Figure~\ref{fig:CoReCo} represent this goal. We expect that the design of such infrastructure can be informed by similar initiatives in different domains. For example, DKPro~\cite{eckartdecastilho2014} provides a collection of reusable Natural Language Processing (NLP) software, enabled by a type system that creates an abstraction layer on top of otherwise incompatible tools, data formats, and models. Similarly, the WEKA platform~\cite{hall2009weka} integrates various machine learning and data mining tools. No such infrastructure exists yet for software engineering analytics.

\emph{Data integration, synthesis, and visualization.}
We know the information needs of code reviewers~\cite{pascarella2018information}. We envision how those information needs can be satisfied to improve code understanding and enable decision-making through an integrated environment. It is still open to see which data is useful and how it can be synthesized to support such decisions. Furthermore, to the best of our knowledge, there is no mechanism to describe context in a way that allows one to define the scope and depth of information necessary to make a decision in code reviews. Work on development task context~\cite{kersten2006using, gasparic2017context} can serve as a starting point here.

\emph{Code review decision making.}
The aggregation of data to enrich the decision-making process in code reviews can, however, also make the problem of analysis paralysis more prominent. The more criteria reviewers have to weigh in their decision, the more complex the solution-finding process becomes.
For example, let's assume a code change intends to improve execution efficiency. Static code analysis of the changed code indicates that code readability is also decreased and below an acceptable threshold. In addition, the code change is located in a feature that has high business value (faster execution will lead to more revenue) but has proven to be defect-prone in the past. What does the reviewer decide? It is unclear to what extent contextualization can alleviate information overload and how it should be implemented to be effective.  








\section{Conclusions}\label{sec:conclusion}
This paper argues for a new generation of code review environments that provide developers with contextualized information to perform reviews. We outline the components of the Code Review Contextualizer (CoReCo) and describe application scenarios that illustrate how this environment would be able to address the information needs of code reviewers. Finally, we outline three research directions that may lead, if followed, to realizing our vision. We hope this work can inspire further in development environments, supporting improvements in developers' experience potentially leading to code quality and development speed gains.

\section*{Acknowledgments}
This work has been supported by ELLIIT; the Strategic Research Area within IT and Mobile Communications, funded by the Swedish Government. The work has also been supported by a research grant for the GIST (reference number 20220235) and SERT project from the Knowledge Foundation in Sweden.

\bibliographystyle{ACM-Reference-Format}
\bibliography{references}

\end{document}